\documentclass[useAMS,usenatbib]{mn2e}
\usepackage{graphicx}
\usepackage{upgreek}
\usepackage{threeparttable}

\def\SNR{\mbox{{SNR~J0533--7202}}}
\newcommand{\D}{$^\circ$}
\newcommand{\SII}{[S\,{\sc ii}]}
\newcommand{\OIII}{[O\,{\sc iii}]}

\title[MULTIFREQUENCY STUDY OF A NEW LMC SNR J0533-7202]
  {Multifrequency study of \SNR, a new supernova remnant in the LMC}
\author[L. M. Bozzetto et al.]
  {L. M.~Bozzetto,$^1$
  M. D.~Filipovi\'c,$^1$ E.J.~Crawford,$^1$ M.~Sasaki,$^2$ P.~Maggi,$^3$
  \newauthor % starts a new line in the
  F.~Haberl,$^3$ D.~Uro\v{s}evi\'c,$^{4,5}$ J. L.~Payne,$^1$ A.~Y.~De~Horta,$^1$ M.~Stupar,$^{6,7}$ 
  \newauthor R. Gruendl,$^8$ \& J.~Dickel$^9$\\
  $^1$School of Computing and Mathematics, University of Western Sydney
    \break Locked Bag 1797, Penrith South DC, NSW 1797, Australia\\
  $^2$Institut f\"ur Astronomie und Astrophysik T\"ubingen, Sand 1, D-72076 T\"ubingen, Germany\\
  $^3$Max-Planck-Institut f\"{u}r extraterrestrische Physik, Giessenbachstra\ss e, D-85748 Garching, Germany\\
  $^4$Department of Astronomy, Faculty of Mathematics, University of Belgrade, Studentski trg 16, 11000 Belgrade, Serbia\\
  $^5$Isaac Newton Institute of Chile, Yugoslavia Branch\\
  $^6$Department of Physics, Macquarie University, Sydney, NSW 2109, Australia\\
  $^7$Australian Astronomical Observatory, PO Box 296, Epping, NSW 1710, Australia\\
  $^8$Department of Astronomy, University of Illinois, 1002 West Green Street, Urbana, IL 61801, USA\\
  $^9$Physics and Astronomy Department, University of New Mexico, MSC 07-4220, Albuquerque, NM 87131, USA
  }
\date{Released 2011 Xxxxx XX}

\pagerange{\pageref{firstpage}--\pageref{lastpage}} \pubyear{2002}

\def\LaTeX{L\kern-.36em\raise.3ex\hbox{a}\kern-.15em
    T\kern-.1667em\lower.7ex\hbox{E}\kern-.125emX}

\begin{document}

\label{firstpage}

\maketitle

\begin{abstract}

We present a detailed study of Australia Telescope Compact Array (ATCA) observations of a newly discovered Large Magellanic Cloud (LMC) supernova remnant (SNR), \SNR. This object follows a horseshoe morphology, with a size 37~pc $\times$ 28~pc (1-pc uncertainty in each direction). It exhibits a radio spectrum with the intrinsic synchrotron spectral index of $\alpha$=--0.47$\pm$0.06 between 73 and 6~cm. We report detections of regions showing moderately high fractional polarisation at 6~cm, with a peak value of 36$\pm$6\% and a mean fractional polarisation of 12$\pm$7\%. We also estimate an average rotation measure across the remnant of --591~rad~m$^{-2}$. The current lack of deep X-ray observation precludes any conclusion about high-energy emission from the remnant. The association with an old stellar population favours a thermonuclear supernova origin of the remnant. 
\end{abstract}


\section{Introduction}

Supernova remnants (SNRs) are responsible for the distribution of heavy elements in the universe, influencing the chemical composition of the next generation of stars. SNRs have a significant effect on their environment, heating up the surrounding gas and dust as the shock waves of the supernova explosion pass through. In turn however, SNRs are heavily impacted by their environment, as their evolution, structure and expansion is greatly affected by the density of the surrounding ISM. In the radio-continuum, SNR emission is predominately non-thermal and will typically display a radio spectral index of $\alpha$ $\sim$ --0.5 (defined by $S \propto \nu^\alpha$). However, this may significantly vary due to environmental factors, different stages of evolution and various types of SNRs.

The Large Magellanic Cloud (LMC) is an irregular dwarf galaxy in close proximity to our own Milky Way galaxy at a distance of 50~kpc \citep{2006ApJ...652.1133M}. As a result of this relatively close distance, objects in the LMC can be observed with significantly better spatial resolution and sensitivity than those in other galaxies. However, it is far away enough that we are able to assume all objects that lie within the galaxy are at very similar distances, therefore making estimates for various analysis methodologies such as extent and surface brightness more accurate, as these values depend on accurate distances to the object. In contrast, objects in the Milky Way can be hard to find accurate distances for, and therefore leads to imprecise measurements. The LMC is also a desirable environment for various astronomical studies as it contains some of the most active star forming regions in our Local Group of galaxies as well as residing outside of the Galactic plane, at a moderate inclination angle of 35$^\circ$ \citep{2001AJ....122.1807V}, which greatly minimises the interference from stars, gas, and dust within the LMC.

%The Australian Telescope Compact Array (ATCA) is a radio interferometer located in northern New South Wales (NSW), Australia. Its new 2~GHz bandwidth coupled with improved sensitivity allows us to observe a category of SNRs that is currently lacking -- the older and more evolved SNRs about to dissipate back into the surrounding ambient ISM. New observations will give us a more complete population to work with and also show through analysis, the final stages of these cosmic blasts.

In this paper, we report on a newly discovered LMC SNR, \SNR.  Radio observations of this object were also taken by \citet{1995A&AS..111..311F,1998A&AS..130..421F,1998A&AS..127..119F} in their surveys of the Magellanic Clouds, however, they did not classify this object. The new observations, data reduction and imaging techniques are described in Section~2. The astrophysical interpretation of newly obtained moderate-resolution total intensity and polarimetric images in combination with the existing Magellanic Cloud Emission Line Survey (MCELS) images are discussed in Section~3.

\section{Observations and Data Reduction}

We observed \SNR\ with the ATCA on the 15$^\mathrm{th}$ and 16$^\mathrm{th}$ of November 2011, using the new Compact Array Broadband Backend (CABB) at array configuration EW367 and at wavelengths of 3 and 6~cm ($\nu$=9000 and 5500~MHz). Baselines formed with the $6^\mathrm{th}$ ATCA antenna were excluded, as the other five antennas were arranged in a compact configuration. The observations were carried out in the so called ``snap-shot'' mode, totaling $\sim$50 minutes of integration over a 14 hour period. PKS~B1934-638 was used for flux density calibration\footnote{Flux densities were assumed to be 5.098 Jy at 6~cm and 2.736 at 3~cm.}  and PKS~B0530-727 was used for secondary (phase) calibration. The phase calibrator was observed twice every hour for a total 78 minutes over the whole observing session. The \textsc{miriad}\footnote{http://www.atnf.csiro.au/computing/software/miriad/}  \citep{1995ASPC...77..433S} and \textsc{karma} \citep{1995ASPC...77..144G} software packages were used for reduction and analysis. More information on the observing procedure and other sources observed in this project can be found in \citet{2012MNRAS.420.2588B, 2012RMxAA..48...41B} and \citet{2012A&A...540A..25D}.

The CABB 2~GHz bandwidth is a 16 times improvement from the previous 128~MHz, and with the new higher data sampling has increased the sensitivity of the ATCA by a factor of 4. The 2~GHz bandwidth not only aids in high sensitivity observations, but also allows data to be split into channels which can then be used for measuring Faraday rotation across the entire bandwidth, at frequencies close enough that the {\it n}~$\times$~180\D ambiguities prevalent when making an estimate between distant frequencies, are no longer an issue.

Images were formed using \textsc{miriad} multi-frequency synthesis \citep{1994A&AS..108..585S} and natural weighting. They were deconvolved with primary beam correction applied. The same procedure was used for both \textit{U} and \textit{Q} stokes parameter maps. 

The 3~cm image (Fig.~\ref{3w6}) has a resolution (full width half maximum (FWHM)) of 21.6\arcsec$\times$15.0\arcsec\ (PA=47.2\D). Similarly, we made an image of \SNR\ at 6~cm (seen as contours in Fig.~\ref{3w6}) which has a FWHM of 34.1\arcsec$\times$26.1\arcsec\ at PA=45.5\D\ and an estimated r.m.s. noise of 0.3 mJy/beam.

%The 6~cm image (seen as contours in Fig.~1) has a resolution of 34.1\arcsec$\times$26.1\arcsec\ at PA=45.5\D\ and an estimated r.m.s. noise of 0.3 mJy/beam. Similarly, we made an image of \SNR\ at 3~cm (Fig.~1) with resolution of 22.0\arcsec$\times$15.0\arcsec\ (PA=47.2\D).

%In addition to our own observations, we made use of a 36~cm Molonglo Synthesis Telescope (MOST) mosaic image (as described in \citealt{1984AuJPh..37..321M}), a 36~cm SUMMS mosaic image \citep{1999AJ....117.1578B}, a 20~cm mosaic image \citep{2007MNRAS.382..543H}, 6 \& 3~cm mosaic images \citep{2010AJ....140.1567D} and lastly 6 \& 3~cm ATCA mosaic visibility files \citep{2005AJ....129..790D}.

%We also used the Magellanic Cloud Emission Line Survey (MCELS) that was carried out with the 0.6~m University of Michigan/CTIO Curtis Schmidt telescope, equipped with a SITE $2048 \times 2048$\ CCD, which gave a field of 1.35\degr\ at a scale of 2.4\arcsec\,pixel$^{-1}$. Both the LMC and SMC were mapped in narrow bands corresponding to \Halpha, \OIII\ ($\lambda$=5007\,\AA), and \SII\ ($\lambda$=6716,\,6731\,\AA), plus matched red and green continuum bands that are used primarily to subtract most of the stars from the images to reveal the full extent of the faint diffuse emission. All the data have been flux-calibrated and assembled into mosaic images. Further details regarding the MCELS are given by Smith et al. (2006) and at http://www.ctio.noao.edu/mcels. Here, for the first time, we present optical images of this object in combination with our new radio-continuum data.

\section{Results and Discussion}

\begin{figure}
\centering\includegraphics[scale=.4, angle=-90]{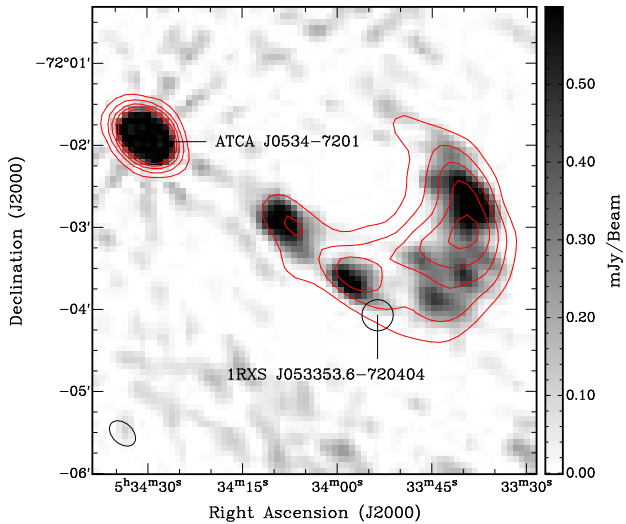}
\caption{ATCA observations of \SNR\ at 3~cm (9000~MHz) overlaid with 6~cm (5500~MHz) contours. The contours are 3, 6, 9, 12 \& 15$\sigma$. The ellipse in the lower left corner represents the synthesised beamwidth (at 6~cm) of 34.1\,\arcsec$\times$26.1\arcsec. The sidebar quantifies the pixel map.  The overlaid circle shows the position of a weak X-ray source seen by ROSAT.
 \label{3w6}}
\end{figure}

\begin{figure}
\centering\includegraphics[trim=0pt 40pt 0pt 0pt,scale=.35]{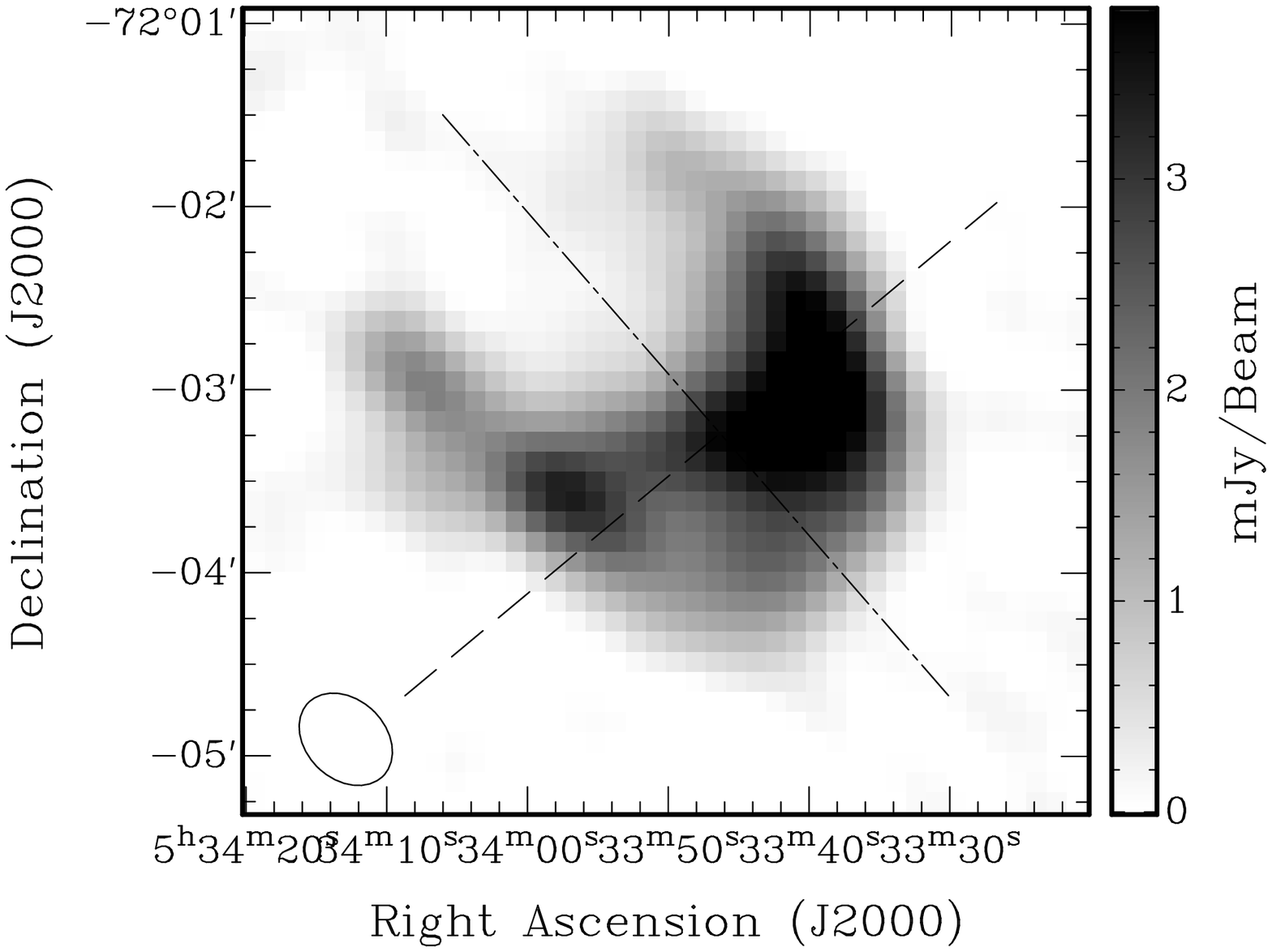}
\centering\includegraphics[angle=-90,scale=.225]{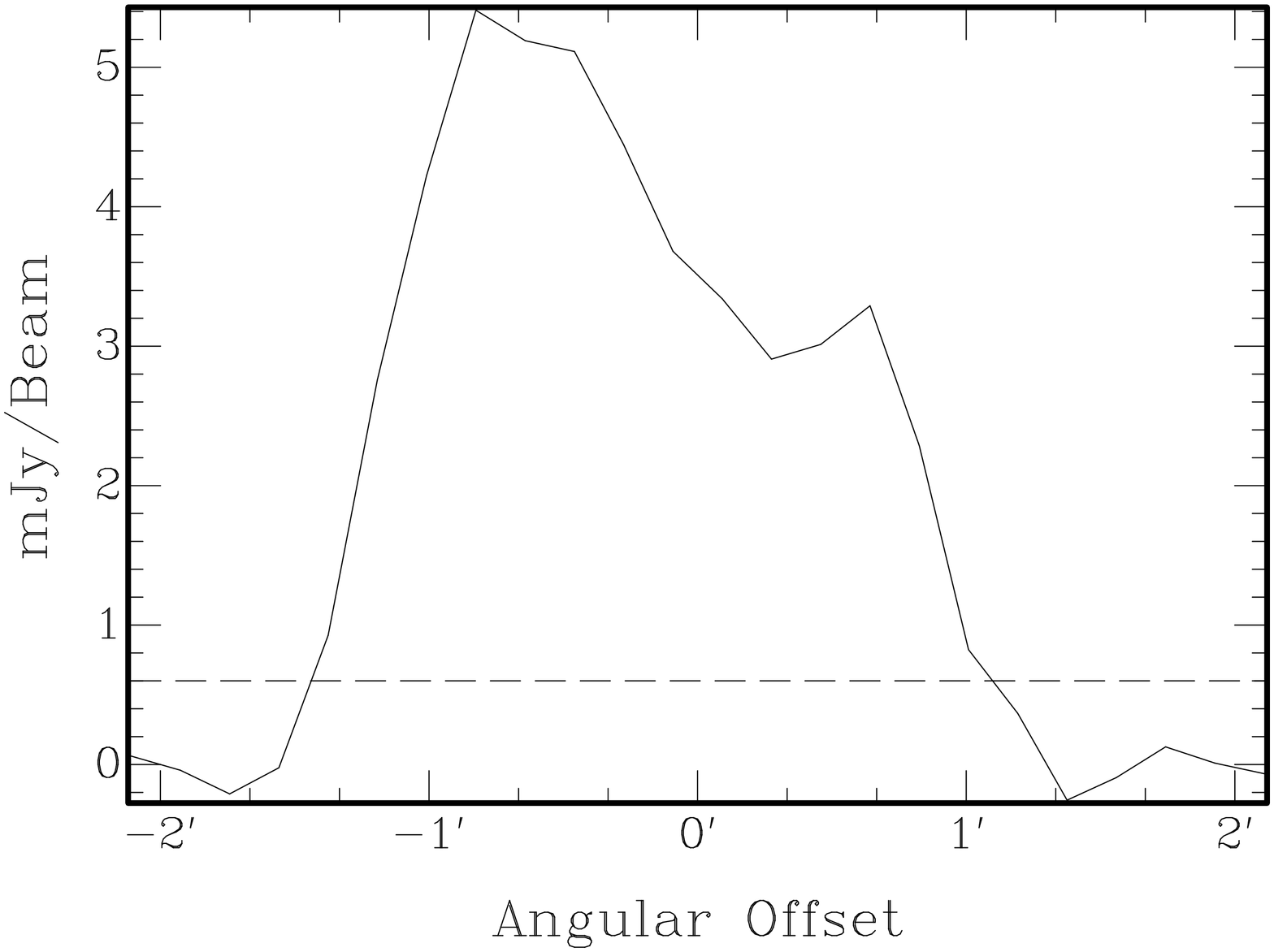}
\centering\includegraphics[angle=-90,scale=.225]{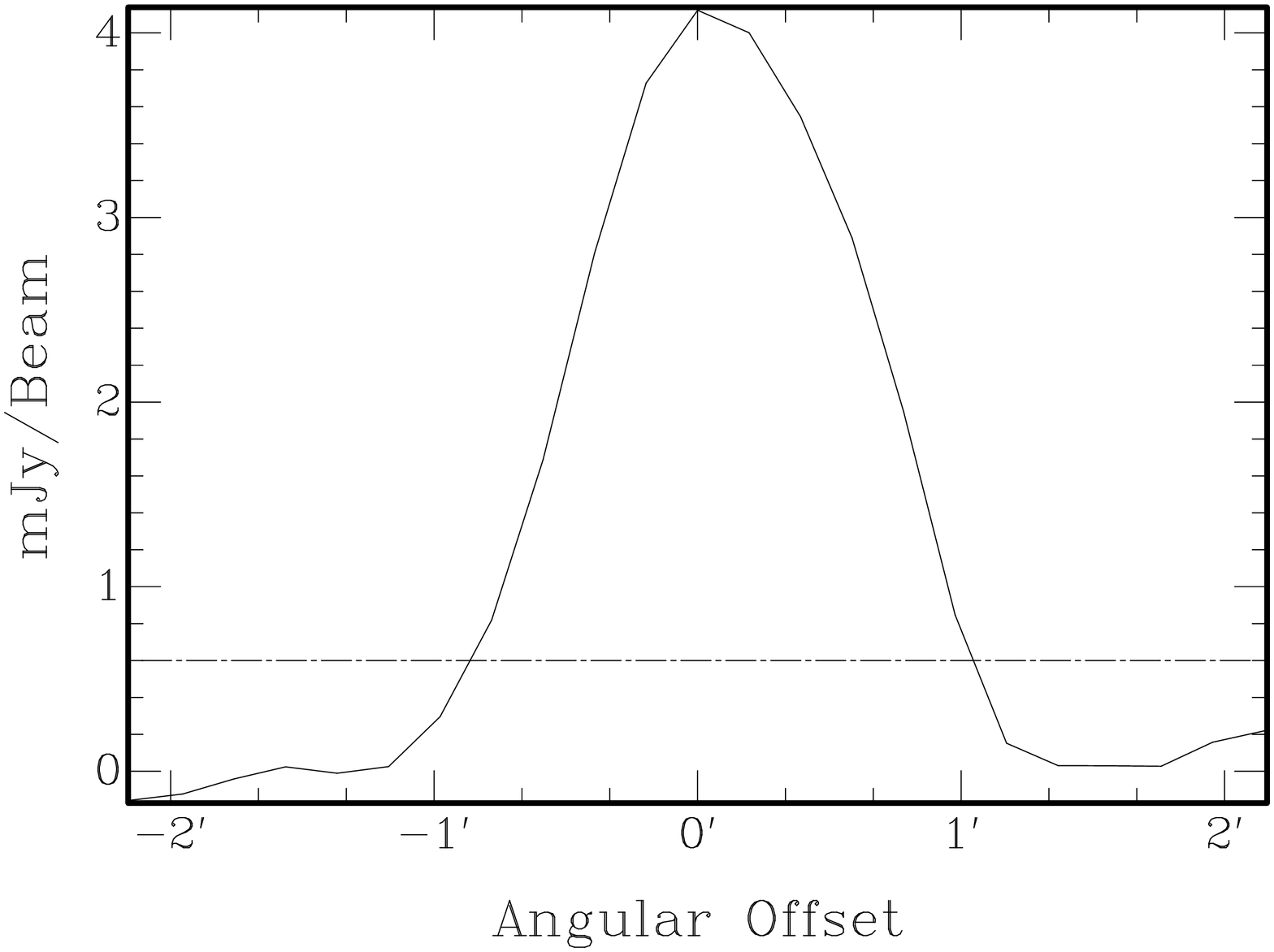}
\caption{The top image shows the 6~cm intensity image of overlaid with the approximate major (NW--SE) and minor (NE--SW) axis. The middle and lower images show the 1-dimensional cross-section along the overlaid lines in the top image, with a superimposed line at 3$\sigma$. 
 \label{extent}}
\end{figure}

This new LMC remnant exhibits a horseshoe morphology (Fig.~\ref{3w6}), centered at RA(J2000)=5$^h$33$^m$46.5$^s$, DEC(J2000)=--72\degr02\arcmin59\arcsec. We selected a one-dimensional intensity profile across the approximate major (NW--SE) and minor (NE--SW) axis (PA=45\D) (Fig.~\ref{extent}) at the 3$\sigma$ noise level (0.9~mJy) to estimate the spatial extent of \SNR. Its size at 6~cm is 152\arcsec$\times$115\arcsec\ with a 4\arcsec\ uncertainty in each direction (37$\times$28 pc with a 1~pc uncertainty in each direction). We did not detect any \OIII\ or \SII\ emission in the Magellanic Cloud Emission Line Survey (MCELS) \citep{2006NOAONL.85..6S} images. However, there is some tentative and very faint H$\alpha$ emission possibly associated with this SNR which wasn't evident anywhere else in the field surrounding the remnant. Although, more sensitive observations are needed.
 
An X-ray source at the rim of the SNR was detected in the course of the ROSAT all-sky survey and was given the identifier 1RXS J053353.6-720404 \citep{1999A&A...349..389V}. However, the likelihood of existence\footnote{In the ROSAT source detection a likelihood L is associated to a probability of a detected source being real P by P = 1 - exp(-L).} in these observations was only 7, meaning only a 3.3$\sigma$ detection, and therefore not much could be done with respect to the X-ray analysis, apart from plotting the ROSAT position (Fig.~\ref{3w6}). This ROSAT position is slightly south-west of one of the radio-bright regions of the remnant. Low statistics preclude any classification between an extended or compact source.

%In the ROSAT source detection a likelihood L was associated to a probability of a detected source being real P by P = 1 - exp(-L). Hence Frank translated this likelihood of 7 to P=99.91\%, which gives the 3.3 sigma detection we quote (so I don't really understand what the reviewer is complaining about...). Maybe you can explicitly gives the formula and/or the probability one can derive from L? For the extent of the source I guess the low statistics preclude any classification between "extended" or "compact" source.
%

%A proposal for future X-ray observations on this SNR using the XMM--Newton was accepted. Hence, an in-depth analysis of the X-ray properties from this remnant will be carried out in the near future.

%Hardness ratios (HR) were taken from ROSAT observations of HR1 = 0.78$\pm$--0.47 and HR2 = --0.65$\pm$--0.27. These values are consistent with an SNR, and might also suggest an older and more evolved remnant. For such an object, we don't expect non-thermal X-ray emission, since there are no non-thermal electrons of that high energies at this age.

This SNR did not appear in the Spitzer mosaic images of the LMC \citep{2006AJ....132.2268M}, neither at the 3.6, 4.5, 5.8, and 8 $\mu$m wavelengths of the IRAC instrument \citep{2004ApJS..154...10F}, nor in the 24, 70, and 160 $\mu$m bands of the MIPS instrument \citep{2004ApJS..154...25R}, suggesting that there is no association with mid or near-infrared wavelengths. There are no OB star candidates in the Magellanic Clouds Photometric Survey (MCPS) catalogue \citep{2004AJ....128.1606Z} within a 100~pc radius around the centre of the remnant, and the star formation history map of the LMC \citep{2009AJ....138.1243H} shows no recent episode of star formation activity in the neighbourhood. The association of SNR J0533-7202 with an old stellar population favours a thermonuclear supernova origin of the remnant.

\begin{figure}
\centering\includegraphics[trim=50 0 0 0, angle=0,scale=.9]{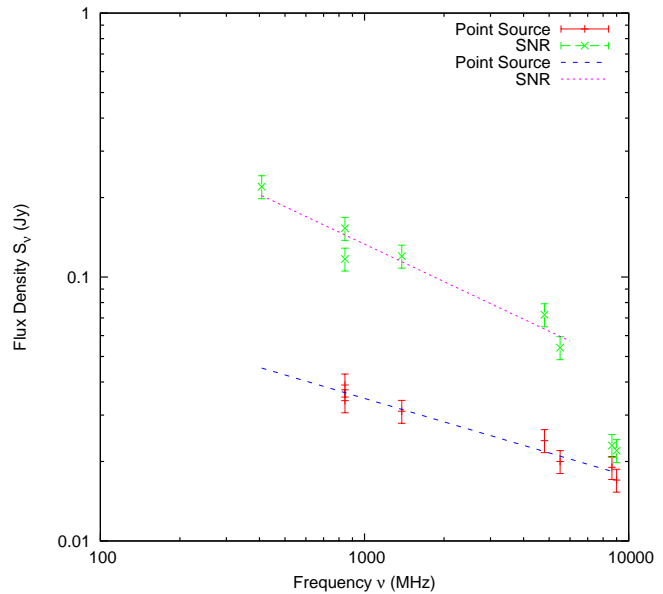}
\caption{Radio-continuum spectrum of \SNR\ and the point source ATCA J0534--7201. The pink { dotted} line represents the spectral index of the SNR and the blue { dashed} line represents the spectrum of the point source.
\label{spcidx}}
\end{figure}

We based the spectral energy distribution (SED) on our own integrated flux estimates, coupled with the 73~cm measurement by \citet{1976AuJPA..40....1C}. These values are shown in Table 1 and then used to produce a spectral index graph (Fig.~\ref{spcidx}). The point source (ATCA J0534--7201; see Fig.~\ref{3w6}) to the east of \SNR\ was unresolved from the SNR in the 73~cm survey (MC4; \citealt{1976AuJPA..40....1C}). We estimate 36~cm flux density measurements from the Molonglo Synthesis Telescope (MOST) mosaic image (as described in \citealt{1984AuJPh..37..321M}) and a 36~cm SUMMS mosaic image \citep{2008yCat.8081....0M}. We point that these values differ some 25\%, which is most likely due to the higher sensitivity of the MOST image and its greater UV coverage. The 20~cm flux density was measured from a mosaic image published by \citet{2007MNRAS.382..543H}. Two different sets of images were used to estimate integrated flux densities at wavelengths of 6 \& 3~cm. The first set (5500 \& 9000~MHz) contains our CABB observations merged with the mosaic visibility files from \citet{2005AJ....129..790D}. These observations use the EW 367 array, which has a shortest baseline of 46~m and as a result, has missing flux from the lack of short spacings. The second set of images (4800 \& 8640~MHz) are from the mosaic images published by Dickel et al. (2010), which used EW 367 and EW 352 arrays at both frequencies as well as Parkes at 4800~MHz but not at 8640~MHz, as the Parkes survey at 3~cm did not extend that far south. This survey data had more spatial frequency coverage, particularly for the extended emission, but less sensitivity. We can see the detrimental effect of missing short spacings in the 3~cm flux density measurements, where they fall well below the trend of the SED at higher wavelengths (Fig.~\ref{spcidx}). Due to the significant impact of the missing shorter spacings (and as a result, missing flux), we omit the 3~cm measurements from the calculation, leaving all the frequencies up to 6~cm (5500~MHz), and thus a spectral index of $\alpha$=--0.47$\pm$0.06.

\begin{figure}
\centering\includegraphics[angle=-90,scale=.38]{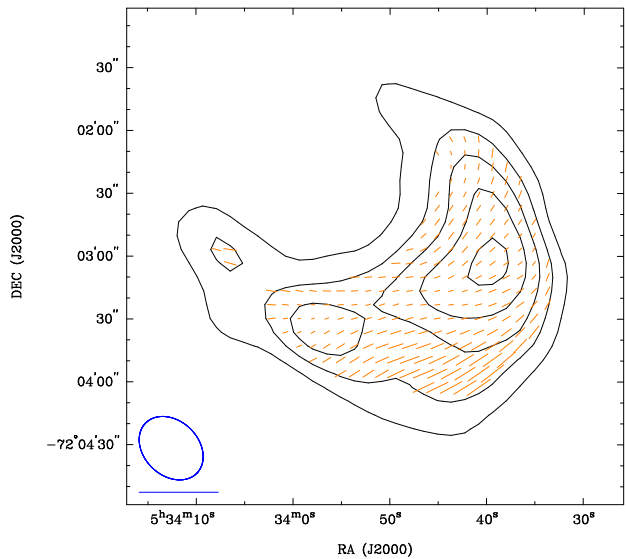}
\caption{ {\textit B}--field vectors overlaid on 6~cm contours (3, 6, 9, 12 and 15$\sigma$) of \SNR\  from ATCA observations. The ellipse in the lower left corner represents the synthesised beamwidth of 34.3\arcsec$\times$26.0\arcsec\ and the line below the ellipse shows a polarisation vector of 100\%.
 \label{polar}}
\end{figure}

Fractional polarisation {(\textit P)} was calculated at 6~cm using:
\begin{equation}
P=\frac{\sqrt{S_{Q}^{2}+S_{U}^{2}}}{S_{I}}
\end{equation}
\noindent where $S_{Q}, S_{U}$ and $S_{I}$ are integrated intensities for \textit{Q}, \textit{U} and \textit{I} Stokes parameters (Fig.~\ref{polar}). Our estimated peak value is 36$\pm$6\% with a mean fractional polarisation of 12$\pm$7\%. This level of fractional polarisation is relatively high when compared to various other SNRs in the LMC \citep{2009SerAJ.179...55C, 2010A&A...518A..35C, 2012MNRAS.420.2588B} and would be (theoretically) expected for an SNR with a radio spectrum of around or less than --$0.5$ \citep{RolfsWilson}. This may indicate varied dynamics along the shell. 

\begin{figure}
\centering\includegraphics[angle=-90,scale=.4]{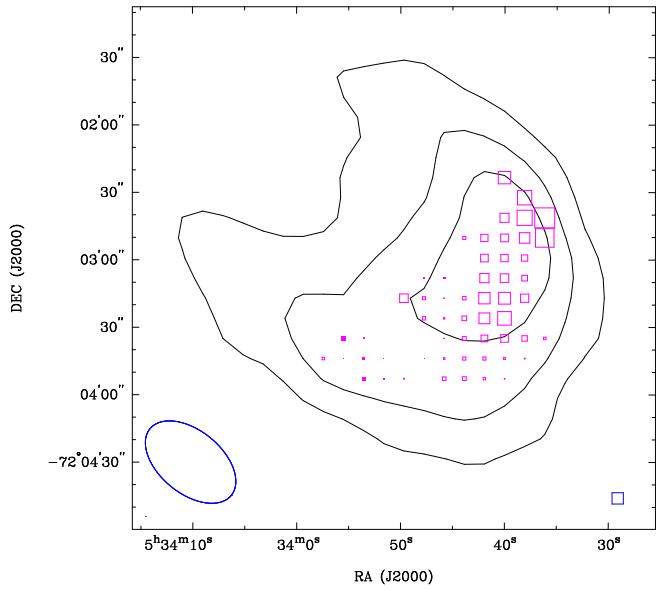}
\caption{Faraday rotation measure of \SNR\ overlaid on 6~cm (128~MHz bandwidth) ATCA contours (3, 8 \& 13$\sigma$). Filled squares represent positive rotation measure while open squares represent negative rotation measure. The ellipse in the lower left corner represents the synthesised beamwidth of 46.7\arcsec$\times$27.9\arcsec\ and the box in the lower right represents a rotation measure of 1000 rad/m$^2$. { The width of the boxes scale with rotation measure.}
 \label{fr}}
\end{figure}

Polarisation position angles were taken from across the 2~GHz bandwidth (at 5500~MHz split into 128~MHz channels) and used to estimate the Faraday rotation for this SNR. The result of this can be seen in Fig.~\ref{fr}, with the open boxes representing negative values of rotation measure and the filled in boxes representing positive rotation measure. The rotation measure varies quite significantly along the SNR with the most change being negative rotation measure near the peak intensity in the western region of the remnant. The average rotation measure across the SNR was --591~rad~m$^{-2}$. This value is nearly double that of the plerion in SNR G326.3--1.8 \citep{2000ApJ...543..840D} and significantly exceeds what is typical of `large' values in the LMC and Milky Way of $\pm$250~rad~m$^{-2}$ \citep{1995AJ....109..200D}. It would be best to treat this value with some level of caution, as at higher radio frequencies -- such as our 6~cm observations -- the amount of rotation measure expected from a SNR is within the same range as the expected error. To mitigate this error and achieve a more reliable value of rotation measure, additional observations would need to be taken, preferably at lower radio frequencies where we would expect a higher level of rotation measure. 

We were also able to estimate the magnetic field strength for this SNR based on the equipartition formula as given in \citet{2012ApJ...746...79A}. This formula is based on the \citet{1978MNRAS.182..443B} diffuse shock acceleration (DSA) theory. By using the spectral index value $\alpha$=--0.5, we get an equipartition value for \SNR\ of $\sim$45 $\upmu$G with an estimated  minimum energy of E$_{min}$ = 9.4$\times$10$^{49}$ ergs.

%A surface brightness to diameter ($\Sigma$ -- D) diagram with the theoretically derived evolutionary tracks by \citet{2004A&A...427..525B} is shown in Fig.~\ref{sb}. \SNR\ is positioned at ({\it D}, $\Sigma$) = (32.5~pc, 6.2~$\times$~10$^{-21}$~W~m$^{-2}$~Hz$^{-1}$~sr$^{-1}$). \SNR\ is likely to be an SNR in the late energy conserving phase, with explosion energy between 0.25 and 1 $\times$ 10$^{51}$ ergs, which evolves in an environment of density $\sim$1 cm$^{-3}$. However, we acknowledge that other possible scenarios are still viable.

From the position of \SNR\ at the surface brightness to diameter ($\Sigma$ - D) diagram ((D, $\Sigma$) = (32.5~pc, 6.2~$\times$~10$^{-21}$~W~m$^{-2}$~Hz$^{-1}$~sr$^{-1}$)) by \citet{2004A&A...427..525B}, we can estimate that \SNR\ is likely to be an SNR in the late energy conserving phase, with an explosion energy between 0.25 and 1 $\times$ 10$^{51}$ ergs, which evolves in an environment of density $\sim$1 cm$^{-3}$.

\section{Conclusion}

We have added a new SNR to the LMC SNR population through conducting a high resolution radio--continuum study of \SNR. We report a relatively large SNR with an extent of $\sim$152\arcsec$\times$115\arcsec\ ($\sim$37$\times$28~pc), and a radio spectral index with $\alpha$=--0.47 between 73 and 6~cm. We estimate fractional polarisation of the remnant at 6~cm with a peak of 36$\pm$6\% and a mean integrated value of 12$\pm$7\%. The lack of recent star formation activity around the remnant makes a thermonuclear supernova origin more likely. \\

\noindent {\bf ACKNOWLEDGMENTS}\\

The Australia Telescope Compact Array is part of the Australia Telescope which is funded by the Commonwealth of Australia for operation as a National Facility managed by CSIRO. We thank the Magellanic Clouds Emission Line Survey (MCELS) team for access to the optical images. This research is supported by the Ministry of Education and Science of the Republic of Serbia through project No. 176005. P.\,M. acknowledges support from the Bundesministerium f\"ur Wirtschaft und Technologie\,/\,Deutsches Zentrum f\"ur Luft- und Raumfahrt (BMWi/DLR) grant FKZ 50 OR 1201.

\begin{table*}
\begin{minipage}{120mm}
 \caption{Integrated flux densities of \SNR\ and the point source ATCA J0534--7201.}
  \begin{threeparttable}
 \label{tbl-1}
 \begin{tabular}{@{}ccccccccc}
  \hline
  $\nu$ & $\lambda$ & R.M.S & Beam Size & S$_{PS}$ & $\Delta$S$_{PS}$ & S$_{SNR}$ & $\Delta$S$_{SNR}$ & Reference\\
  (MHz) & (cm) & (mJy)& (\arcsec) & (mJy) & (mJy) & (mJy) &(mJy) &   \\
  \hline
73 & 408 & 40 & 157.2$\times$171.6 & ---  & --- & 220 & 22 &  \citet{1976AuJPA..40....1C}\\
36$^a$ & 843 & 0.8 & 46.4$\times$43.0 & 34 & 3 & 153 & 15 & This work\\
36$^b$ & 843 & 1.0 & 46.4$\times$43.0 &  39 & 4 & 117 & 12 & This work\\
20 &1384 & 0.6 & 40.0$\times$40.0 & 31 & 3 & 120 & 12 & This work\\
6 & 4800 & 0.7 & 35.0$\times$35.0 & 24 & 2 & 72 & 7 & This work\\
6 & 5500 & 0.3 & 34.1$\times$26.1 & 20 & 2 & 54 & 5 & This work\\
3$^c$ & 8640 & 0.7 & 22.0$\times$22.0 & 19 & 2 & 23 & 2 & This work\\
3$^c$ & 9000 & 0.3 & 21.6$\times$15.0 & 17 & 2 & 22 & 2 & This work\\
  \hline
 \end{tabular}
 \begin{tablenotes}
 \item $^a$ -- Uses the MOST mosaic image
 \item $^b$ -- Uses the SUMMS mosaic image
 \item $^c$ -- Due to short spacing, these 3~cm flux density estimates are omitted from the SNR SED estimate.
 \end{tablenotes}
 \end{threeparttable}
\end{minipage}
\end{table*}

%\begin{figure}
%\centering\includegraphics[scale=.4, angle=-90]{0533-7202-3w6-2}
%\caption{ATCA observations of \SNR\ at 3~cm (9000~MHz) overlaid with 6~cm (5500~MHz) contours. The contours are 3, 6, 9, 12 \& 15$\sigma$. The ellipse in the lower left corner represents the synthesised beamwidth (at 6~cm) of 34.1\,\arcsec$\times$26.1\arcsec. The sidebar quantifies the pixel map.  The overlaid circle shows the position of a weak X-ray source seen by ROSAT.
% \label{3w6}}
%\end{figure}

%\begin{figure}
%\centering\includegraphics[scale=.45]{0533-7202-3cm-wconts}
%\caption{ATCA observations of \SNR\ at 3~cm (9000~MHz) overlaid with 6~cm (5500~MHz) contours. The contours are 3, 9, \& 18$\sigma$. The ellipse in the lower left corner represents the synthesised beamwidth (at 6~cm) of 46.7\,\arcsec$\times$27.9\arcsec. The sidebar quantifies the pixel map and its units are mJy/beam. 
% \label{3w6}}
%\end{figure}

%\begin{figure}
%\centering\includegraphics[trim=0 0 0 50,angle=-90,scale=.35]{0533-mcels-lf}
%\caption{MCELS composite optical image \textrm{(RGB =H$\alpha$,[S\textsc{ii}],[O\textsc{iii}])} of \SNR\ overlaid with 6~cm contours. The contours are 3, 11 and 19$\sigma$.
% \label{mcel}}
%\end{figure}

%\begin{figure}
%\centering\includegraphics[angle=-90,scale=.4]{0533-7202-rm-f}
%\caption{Faraday rotation measure of \SNR\ overlaid on 6~cm intensity image. Filled squares represent positive rotation measure while open squares represent negative rotation measure. 
% \label{fr}}
%\end{figure}

%\begin{figure}
%\centering\includegraphics[scale=1, trim=270 270 300 190]{0533sb}
%\caption{Surface brightness-diameter diagram of SNRs from Berezhko \& Volk (2004), overlaid with a red star representing the position of  \SNR. The evolutionary tracks are for ISM densities of {\it N}$_H$ = 3, 0.3 and 0.003~cm$^{-3}$ and explosion energies of {\it E}$_{\sc SN}$ = 0.25, 1 and 3 $\times$ 10$^{51}$ erg.
%\label{sb}}
%\end{figure}

%\begin{figure}
%\centering\includegraphics[scale=.35, angle=-90, trim=0 0 0 0]{0533-4732}
%\centering\includegraphics[scale=.35, angle=-90, trim=140 0 0 0]{0533-5244} % ? L B ? 
%\centering\includegraphics[scale=.35, angle=-90, trim=140 0 0 0]{0533-5756}
%\centering\includegraphics[scale=.35, angle=-90, trim=140 0 0 0]{0533-6268}
%\caption{Radio-continuum images of \SNR\ taken by splitting the original 2~GHz bandwidth at 6~cm into four 512~MHz bands.
%\label{sb2}}
%\end{figure}

\bibliographystyle{mn2e}
\bibliography{0533-7202-v_4_refs}

\label{lastpage}

\end{document}